\documentclass[twocolumn,showpacs,preprintnumbers,amsmath,amssymb]{revtex4}

\usepackage{graphicx}
\usepackage{dcolumn}
\usepackage{bm}

\begin{document}


\title{Structural, electronic, and magnetic properties of a ferromegnetic semiconductor: Co-doped TiO$_2$ rutile}

\author{W. T. Geng$^{1,2}$}
\author{Kwang S. Kim$^1$}%
\affiliation{%
$^1$National Creative Research Initiative Center for Superfunctional Materials and Department of Chemistry, Pohang University of Science and Technology, Pohang 790-784, Korea \\
$^2$Fritz-Haber-Institut der Max-Planck-Gesellschaft, Faradayweg 4-6, D-14195 Berlin, Germany
}%

\date{\today}

\begin{abstract}
Room-temperature ferromagnetism has been recently discovered in Co-doped  TiO$_2$ rutile. 
Our $ab$ $initio$ density-functional theory investigations show that the substitutional Co ions incorporated into TiO$_2$ rutile tend to cluster and then the neighboring interstitial sites become energetically favorable for Co to reside. 
This suggests that a Co-doped rutile containing only substitutional Co is not an appropriate reference bulk system in derterming the local environment of Co in polycrystalline (Ti,Co)O$_2$ rutile.
We also find that the interstitial Co is in the low spin state and destroys the spin-polarization of the surrounding substitutional Co, hence reduces the average magnetic moment of impurity atoms. 

\end{abstract}

\maketitle
\section{INTRODUCTION}	
Spintronics has emerged expeditiously as a novel technology where both charge and spin are harnessed for new functionalities in devices combining standard microelectronics with spin-dependent effect.\cite{1,2,3}.
For their practical spintronic applications, a ferromagnetic semiconductor is required to be not only easily integerable but also having a high Curie temperature. 
The development of new materials grows rapidly. Room-temperature ferromagnetism has been observed, for instance, in Mn doped ternary compounds such as CdGeP$_2$\cite{CdGeP2}, ZnSnAs$_2$\cite{ZnSnAs2}, and ZnGeP$_2$\cite{ZnGeP2}. 
A recent discovery of room-temperature ferromagnetism in Co-doped TiO$_2$ anatase by  Matsumoto $et$ $al$. using combinatorial molecular beam epitaxy (MBE) technique\cite{science} has motivated intensive studies on the structural and physical properties of this material.\cite{apl,prb,cluster,geng}
Transparent to visible light, this material is expected to be of particular importance in spin-based optoelectronic applications such as spin-light-emission-diode and optical switches operating at terahertz frequency\cite{3}.

It is interesting to note that Matsumoto $et$ $al$. observed magnetic structure only in (Ti,Co)O$_2$ anatase but not in rutile phase\cite{ass}, seemingly suggesting that the magnetization of the impurity atoms depends sensitively on the local environment surrounding them.
More recently, Park  $et$ $al$. have successfully grown ferromagnetic  (Ti,Co)O$_2$ rutile films by sputtering\cite{sputter}.
The Curie temperature was estimated to be above 400 K for Co content of 12$\%$. Since the rutile phase is thermodynamically more stable than the anatase phase, it may have higher potential for technical applications. The dissimilar behaviors of films grown by different methods indicate that the local environment of Co could have drastic effect on the magnetic structure of this material. Furthermore, by annealing the (Ti,Co)O$_2$ samples prepared by the sol-gel method at different temperatures, Soo $et$ $al$. found that the local structure arround Co could remain virtually anatase-like while the nanocrystalline materials underwent anatase-to-rutile phase transition\cite{Soo}. This phenomenon is rather unique and certainly deserves more detailed investigation in order to gain a microscopic understanding of the magnetism displayed by this intriguing material.

We here point out that in extracting the information of the local structure of Co through the X-ray absorption fine structure method, Soo  $et$ $al$. employed a refernce bulk system assuming all Co ions substitute for the Ti sites in rutile TiO$_2$\cite{Soo}. In a recent work, we showed by $ab$ $initio$ density functional theory study that Co could also occupy the interstitial sites in anatase  TiO$_2$. The question arise is: Is it also the case for Co-doped rutile  TiO$_2$?
To answer this question, we have performed a computational investigation on the structural, electronic, and magnetic properties of Co-doped TiO$_2$ rutile. 
We show that the substitutional Co ions tend to cluster, but this has no remarkable effect on the magnetization of Co. We also find that an interstitial site will become energetically favorable for a Co atom (in reference to bulk Co) when two of its neighboring substitutional sites are occupied by Co. 
The interstitial Co is in the low spin state (0.94$\mu_B$), but it exerts strong detrimental effect on the magnetization of its neighboring Co (0.76$\mu_B$ $\rightarrow$ $-0.03\mu_B$) and therefore reduces the magnetic moment Co in this magnetic semiconductor. 
In the following, we give a brief outline of our theoretical procedure, a discussion of the results based on bonding characteristics and its relevance to our understanding of the magnetism in this Co-doped oxide.

The calculations were done using a spin-polarized version of the Vienna {\it ab initio} Simulation Package (VASP)\cite{vasp1}. Fully nonlocal Vanderbilt-type ultrasoft pseudopotentials\cite{pp} were used to represent the electron-ion interaction. Exchange correlation interactions were described by the Perdew-Wang 1991 generalized gradient approximation (GGA)\cite{pw91}. 
Wave functions and  augmentation charges were expanded in terms of plane waves with an energy cutoff of 396 and 700 eV, respectively. 
For each system, the geometry was relaxed until the atomic forces were smaller than 0.03 eV/\AA.

The 6-atom rutile unit cell of TiO$_2$ (see Figure 1a) is characterized by the two lattice constants $a$ and $c$ and the internal parameter $u$. The positions of Ti are (0, 0, 0) and (1/2, 1/2, 1/2); and those of O are ($\pm u$, $\pm u$, 0) and (1/2$\pm u$, 1/2$\mp u$, 1/2). The optimized $a$, $c$, and $u$ are 4.661 \AA, 2.973 \AA, and 0.638, in good agreement with experiment\cite{wyckoff} and previous pseudopotential GGA calculations\cite{rutile}.
For Co-doped cases, we mainly dealt with a supercell containing 16 primitive unit cells ($2a\times 2a\times 2c$, see Figure 1b) with one or two Ti atoms (site B1, or, sites B1 and B2) replaced by Co and with/without an interstitial Co (site A). 
Base on our knowledge of anatase (Ti,Co)O$_2$ that volume contraction and expansion associated with substitutional and interstitial Co have only negligible effect on the formation heat of  (Ti,Co)O$_2$\cite{geng}, we fixed $a$ and $c$ at the calculated bulk TiO$_2$ values and optimized only internal freedoms upon Co-doping.

To see whether substitutional Co (Co$^S$) cluster or not, we first minimized the free energy of the cell Ti$_{7}$CoO$_{16}$ (Co content: 12.5$\%$). 
Then we doubled this cell and brought the two Co$^S$ together to form Ti$_{14}$Co$^S_{B1}$Co$^S_{B2}$O$_{32}$ ($2a\times 2a\times 2c$). 
We find the latter configuration is more stable by 0.31eV/Co, indicating that Co clustering does occur at this concentration.  
The calculated formation energy of an interstitial Co (Co$^I$) as a neutral defect in reference to bulk Co, $E$(Ti$_{16}$Co$^I_A$O$_{32}$)-$E$(Ti$_{16}$O$_{32}$)-$E$(Co), is +1.75 eV, implying a sole Co$^I$ is highly unstable in rutile TiO$_2$. This can be understood in the bond-order$-$band-strength picture. Ti:O + Co:O is much weaker than O:Ti:O, as is evident from the comparison of the formation heat of TiO$_2$, TiO, and CoO\cite{CRC}.
 Nonetheless, with the presence of Co$^S_{B1}$, the formation energy of Co$^I$, i.e., $E$(Ti$_{15}$Co$^S_{B1}$Co$^I_A$O$_{32}$)-$E$(Ti$_{15}$Co$^S_{B1}$O$_{32}$)-$E$(Co), drops to +0.47 eV. According to our calculations, the formation heat of CoO$_2$ in a hypothetical rutile TiO$_2$ structure is $-3.46$ eV, slightly higher than two times of that of CoO ($-1.59$ eV). This means that  Co:O + Co:O is only slightly weaker than O:Co:O , thus Co$^I$ becomes much less unstable.  Further calculations show that when both B1 and B2 site are occupied by Co upon clustering, site A becomes favorable for a Co$^I$ with a formation energy of $-0.72$ eV. Once again, this can be explained by the higher formation heat of Co$_3$O$_4$ ($-9.57$ eV) than two times of that of CoO$_2$ ($-3.46$ eV).
Our calculations thus reveal a 2Co$^S$+Co$^I$ local structure in Co-doped TiO$_2$ rutile.
It should be noted that the formation of more complicated (Co$^S$, Co$^I$) combinations is also possible depending on the growth condition, but the formation probability should be lower than  2Co$^S$+Co$^I$. 

In order to examine the effect of clustering on magnetism, we performed calculations on both intra-cell FM and intra-cell AFM alignments of the Co pair in a Ti$_{14}$Co$^S_{B1}$Co$^S_{B2}$O$_{32}$ unit cell (For simplicity, we denote this cell as Ti$_{14}$Co$_2$O$_{32}$ in the following) while keeping the inter-cell coupling ferromagnetic. 
The AFM phase is found to be unstable and it switches to FM during self-consistency iterations. When the two Co ions are far apart by 7.23 \AA with one at (0, 0, 0) and the other at ($a$, $a$, $c$), the FM phase is lower in totl energy by 0.26 eV/Co than the AFM phase. In the case of 2Co$^S$+Co$^I$, we checked all the three possible intra-cell spin alignments of the three Co ions. All of them converge to the same final state, with both Co$^I$ and Co$^S$ in the low spin state. 
The calculated total spin magnetic moment in a unit cell of Ti$_{15}$Co$_1$O$_{32}$, Ti$_{14}$Co$_2$O$_{32}$, and Ti$_{14}$Co$_3$O$_{32}$, and that in the atomic sphere of Co (radius=1.10 \AA ) are listed in Table I. 
It is seen that in all three cases, Co ions are in the low spin state, regardless to their oxidation states. The spin moment of Co$^S$ does not change much upon clustering. Co$^I$, on the other hand, behaves rather differently. When coupled with Co$^S$, Co$^I$ kills almost entirely the spin moments of  Co$^S$. As a result, the average spin magnetic moment of Co is reduced to 0.29 $\mu_B$, less than a half of the non-Co$^I$ case. 

Figure 2 displays the calculated density of states for non-doped and various cases of Co-doped rutile. Like in anatase\cite{prb,geng}, the Co states are mainly located in the energy-gap region and the O and Ti states are not much affected by Co doping. It is worth noting that with GGA, while  Ti$_{15}$Co$_1$O$_{32}$ and Ti$_{14}$Co$_2$O$_{32}$ just miss being insulators, the Fermi energy of their rutile counterparts just overpass the valence bands and fall into the energy gap. The presence of  Co$^I$ does not change the insulating electronic structure of this material although sole Co$^I$ ions would turn the material into a metallic system (DOS not shown), and as a consequence the magnetic interaction between Co ions at low temperature remains to be superexchange in the absence of carriers. 

In Figure 3, we compare the $3d$ DOS of Co in rutile with with those of Co in other Co oxides with different oxidation states such as CoO$_2$ (+4) and Co$_3$O$_4$ (+3 and +2). CoO$_2$, the end member of LiCoO$_2$, has recently been isolated\cite{coo2_1}. It seems to be of CdCl$_2$ type with a monoclinic distortion, but positions of O remain to be determined. In this regard, we believe a study on a hypothetical CoO$_2$ in TiO$_2$ rutile structure is meaningful in illustrating the electronic structure of Co
$^{4+}$. For both  CoO$_2$ and Co$_3$O$_4$, we have optimized their internal structural parameters.
We can see that although much more localized, $d$ DOS of Co$^S$ in Ti$_{15}$Co$_1$O$_{32}$ resembles that of Co in  CoO$_2$ in both exchange splitting and crystal field splitting, suggesting they are in +4 oxidation state. In an octahedral field, $t_{2g}$ is lower than $e_g$ and thus Co$^{4+}$ will have a spin magnetic moment of 1$\mu_B$. 
As for the combination  2Co$^S$+Co$^I$, a close resemblance can be found between Co$^S$ (Co$^I$) in TiO$_2$ rutile and Co$^B$ (Co$^A$) in Co$_3$O$_4$. It thus provides a strong evidence that  Co$^S$ and Co$^I$ in such a combination are in +3 and +2 oxidation states, respectively. The six $d$ electrons of Co$^S$ just fill  $t_{2g}$ states and thus yield a vanishing spin magnetic moment. Unlike the interstial Co in anatase and Co$^A$ in Co$_3$O$_4$, Co$^I$ in  TiO$_2$ rutile is in an octahedral field, therefore the lone electron in  $e_g$ state gives a spin moment of 1$\mu_B$. 

In summary, our $ab$ $initio$ density functional theory study reveals that in Co-doped rutile TiO$_2$, Co can occupy interstitial sites, upon clustering of sbstitutional Co. As a result, a reference bulk system used to extract information of the local structure of Co in a polycrystalline material should be one that containing both substitutional and interstitial Co. It is found that the interstitial Co destroys the spin-polarization of the neighboring substitutional impurities by lowering its charge state and therefore plays an important role in the magnetic structure of such material. 
     
\begin{acknowledgments}
Work in Korea was supported by Korea Institute of Science and Technology Evaluation and Planning (Creative Research Initiative) and in Germany by Max-Planck-Society Fellowship. W.T.G. thanks Dr. Qiang Sun for helpful discussions.
\end{acknowledgments}

 \begin{figure}
 \caption{\label{fig1} Calculational unit cells of non-doped (panel a) and Co-doped TiO$_2$ rutile (panel b). Large and small circles represent O and Ti (Co), respectively.}
 \end{figure}

 \begin{figure}
 \caption{Total density of states for non-doped and various cases of Co-doped TiO$_2$ rutile. Dotted vertical lines denote the Fermi energy.}
 \end{figure}

 \begin{figure}
 \caption{The calculated density of $d$ states of Co in CoO$_2$, Co$_3$O$_4$, and various cases of Co-doped TiO$_2$ rutile. Dotted vertical lines denote the Fermi energy.\label{fig3}}
 \end{figure}

\begin{table}
\caption{\label{tab:table3}
Calculated spin magnetic moment ($\mu_B$) in the atomic sphere of Co (radius = 1.10 \AA) and that for a whole unit cell. Co(B1), Co(B2),
 and Co(A) denote Co ions occupying sites B1, B2, and A (see Figure 1b). }
\begin{ruledtabular}
\begin{tabular}{ccccc}
Unit cell & Co(B1) & Co(B2) & Co (A) & Total \\ \hline
Ti$_{15}$Co$_1$O$_{ 32}$ & 0.73 & -   & -   & 1.00  \\
Ti$_{14}$Co$_2$O$_{ 32}$ & 0.77 & 0.75 & -  & 2.00  \\
Ti$_{14}$Co$_3$O$_{ 32}$ &-0.03 &-0.03 & 0.94 & 1.00 \\
\end{tabular}
\end{ruledtabular}
\end{table}

\bibliography{ti}

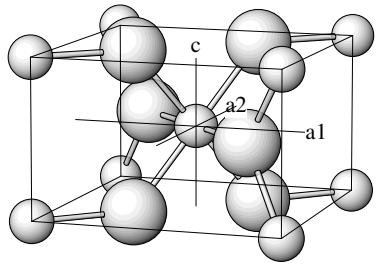
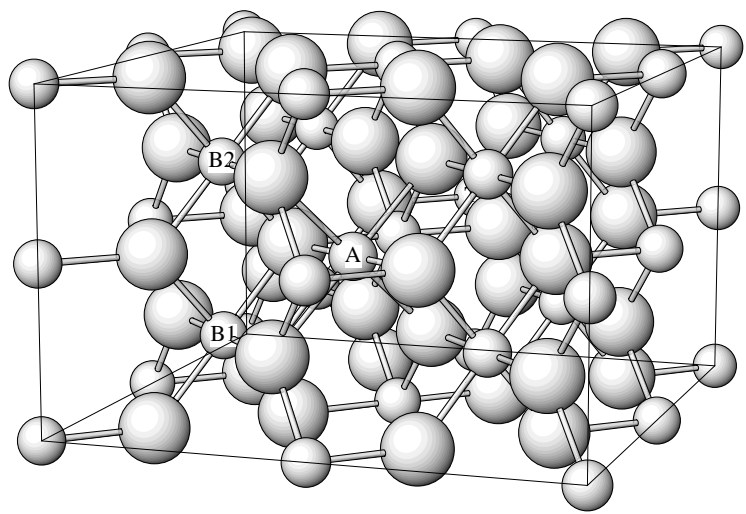
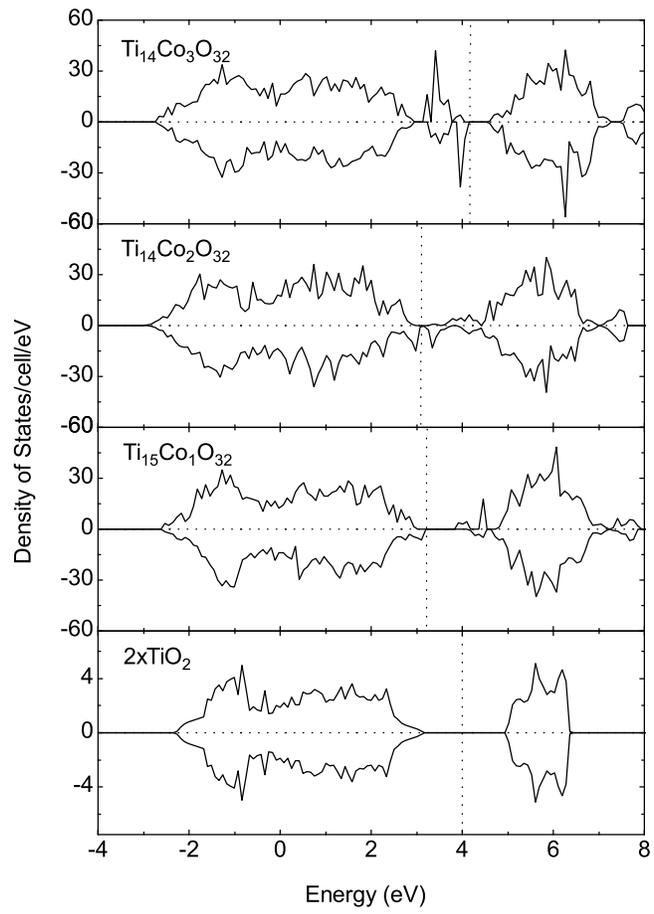
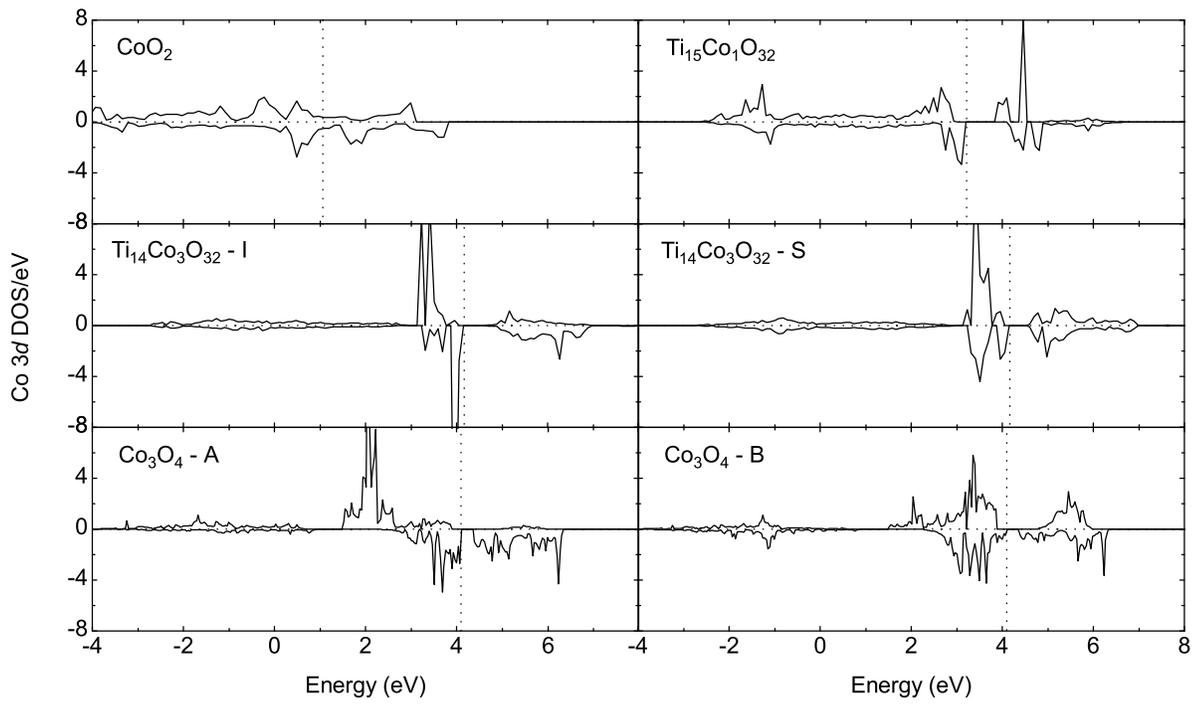
\end{document}